\documentclass[%
 reprint,
 amsmath,amssymb,
 aps,
prb,
]{revtex4-1}

\usepackage{graphicx}
\usepackage{dcolumn}
\usepackage{bm}

\newcommand{\ie}{\textit{i.e.}\/, }
\newcommand{\eg}{\textit{e.g.}\/, }
\newcommand{\cf}{\textit{cf.}\/, }

\providecommand*{\mrm}[1]{\mathrm{#1}}
\providecommand*{\unit}[1]{\ensuremath{\mrm{\,#1}}}
\providecommand*{\eu}{\ensuremath{\mrm{e}}}
\providecommand*{\iu}{\ensuremath{\mrm{i}}}

\renewcommand{\Re}{\ensuremath{\mrm{Re}}}	
\renewcommand{\Im}{\ensuremath{\mrm{Im}}}	

\def\XXint#1#2#3{{\setbox0=\hbox{$#1{#2#3}{\int}$}
     \vcenter{\hbox{$#2#3$}}\kern-.5\wd0}}

\begin{document}


\title{On the quasistatic optimal plasmonic resonances in lossy media}

\author{Sven~Nordebo}
 \email{sven.nordebo@lnu.se}
\affiliation{%
 Department of Physics and Electrical Engineering, Linn\ae us University, 351 95 V\"{a}xj\"{o}, Sweden.
}%

\author{Mohammad~Mirmoosa}
\email{mohammad.mirmoosa@aalto.fi}
\author{Sergei~Tretyakov}
\email{sergei.tretyakov@aalto.fi}
\affiliation{
 Department of Electronics and Nanoengineering, Aalto University, P.O. Box 15500, FI-00076 Aalto, Finland.
}%

\date{\today}

\begin{abstract}
This paper discusses and analyzes the quasistatic optimal plasmonic dipole resonance of a small dielectric particle embedded in a lossy surrounding medium.
The optimal resonance at any given frequency is defined by the complex valued dielectric constant that maximizes the absorption of the particle under the quasistatic approximation 
and a passivity constraint. In particular, for an ellipsoid aligned along the exciting field, the optimal material property is given by the complex conjugate of the 
pole position associated with the polarizability of the particle. In this paper, we employ classical Mie theory to analyze this approximation for spherical 
particles in a lossy surrounding medium. It turns out that the quasistatic optimal plasmonic resonance is valid provided that the electrical 
size of the particle is sufficiently small at the same time as the external losses are sufficiently large. Hence, it is important to note that this approximation can 
not be used for a lossless medium, and which is also obvious since the quasistatic optimal dipole absorption becomes unbounded for this case.
Moreover, it turns out that the optimal normalized absorption cross section area of the small dielectric sphere has a very subtle limiting behavior, 
and is in fact unbounded even in full dynamics when both the electrical size as well as the exterior losses tend to zero at the same time. 
A detailed analysis is carried out to assess the validity of the quasistatic estimation of the optimal resonance and numerical examples are included to illustrate the asymptotic results.
\end{abstract}

\maketitle


\section{Introduction}

The classical theories as well as most of the recent theoretical studies and experiments regarding the efficiency of plasmonic resonances 
reported in the literature are concerned with metal nanoparticles where the exterior domain is lossless, 
see \eg \cite{Bohren+Huffman1983,Link+etal2000,Maier2007,Tretyakov2014,Miller+etal2016,Tzarouchis+etal2016}. 
As \eg in \cite{Miller+etal2016}, a variational approach is employed in connection with a generalized optical theorem for scattering, absorption and extinction 
\cite{Lytle+etal2005}, to obtain an upper bound on the absorption that can be achieved inside a scatterer with arbitrary geometry and with a given volume and material property.
Furthermore, an upper bound on the dipole absorption of an electrically small particle with arbitrary geometry and structural parameters is given in \cite{Tretyakov2014}.
The bound \cite[Eq.~(16) on p.~937]{Tretyakov2014} is based on the 
optical theorem \cite{Bohren+Huffman1983} and obtained by optimizing absorption 
directly in terms of the complex valued polarizability of the particle. 
Both results in \cite{Tretyakov2014} and \cite{Miller+etal2016} are valid for a lossless surrounding medium only. 

However, there are many application areas of plasmonics where the exterior losses must be taken into account.
One potential application in medicine is the localized electrophoretic heating of a bio-targeted and electrically 
charged gold nanoparticle (GNP) suspension as a radiotherapeutic hyperthermia based method to treat cancer, 
\cf \cite{Corr+etal2012,Sassaroli+etal2012,Collins+etal2014,Nordebo+etal2017a,Dalarsson+etal2017a}
or with the related plasmonic photothermal therapy as proposed in \cite{Huang+etal2008}.
Other potential application areas include plasmon waveguides, aperture arrays, extraordinary transmission, superlenses, artificial magnetism, negative refractive index,
and surface-enhanced biological sensing with molecular monolayer spectroscopy, etc., see \eg \cite{Maier2007}. 

There has been a number of investigations devoted to the scattering, absorption and extinction of small particles embedded 
in a lossy medium, see \eg \cite{Mundy+etal1974,Chylek1977,Bohren+Gilra1979,Lebedev+etal1999,Sudiarta+Chylek2001,Durant+etal2007a}. 
This topic has even been subjected to some controversy due to the difficulties to define a general theory
encompassing the notion of a cross section when the surrounding medium is lossy, see \eg\cite{Bohren+Gilra1979,Lebedev+etal1999,Sudiarta+Chylek2001,Durant+etal2007a}. 
In contrast, for a lossless surrounding medium the absorption cross section can be defined
from the power flowing into a conceptual sphere surrounding the particle at an arbitrary radius, and which enables the derivation of an optical theorem valid for arbitrary geometries \cite[pp.~71 and 140]{Bohren+Huffman1983}. For a lossy medium this theory is no longer valid, which is due to the fact
that the absorption in the surrounding medium depends on the geometry of the scatterer.
Hence, the optical theorems for lossy media are typically given only for spheres. As \eg in \cite{Nordebo+etal2018f} is given 
new fundamental upper bounds on the multipole absorption and scattering of a rotationally invariant
sphere embedded in a lossy surrounding medium and which are derived based on the corresponding generalized optical theorem 
as given in \eg \cite[Eq.~(7) on p.~1276]{Sudiarta+Chylek2001}.

We are concerned here with the optimal absorption of an electrically small spherical object 
embedded in a lossy surrounding medium when the near-field distribution can be found using the quasistatic approximation. 
For simplicity, we are considering only the important special case of non-magnetic, dielectric materials which are common in plasmonic applications. 
Magnetic materials can be treated similarly.
We are also considering only a single electric dipole resonance, and we are
discarding any possibilities of having an unbounded absorption due to multiple mode super resolution effects, etc.,
see \eg \cite{Valagiannopoulos+etal2015,Maslovski+etal2016,Valagiannopoulos+Tretyakov2016}.
A quasistatic theory has been developed in \cite{Nordebo+etal2017a,Dalarsson+etal2017a} giving the optimal plasmonic dipole resonance of small dielectric 
ellipsoids in terms of an optimal conjugate match with respect to the background loss. 
However, an important limitation of this theory is that it does not give the correct physical answers when the background becomes lossless or has very small losses.
This limitation is obvious since the quasistatic optimal absorption becomes unbounded in the case when the external losses vanish
\cf \eg \cite[Eq.~(33) on p.~5]{Dalarsson+etal2017a}, 
in contrast to the well known fact that the absorption cross section of a small dipole scatterer in a lossless exterior medium is bounded 
by $3\lambda^2/8\pi$, \cf \cite[Eq.~(16) on p.~937]{Tretyakov2014}.

This limitation, which may even appear as a contradiction, can be understood simply by realizing that the quasistatic model disregards radiation damping. 
Moreover,  the plasmonic singularity of the sphere ($\epsilon=-2$ in vacuum) 
exists only in the sense of a limit as the size of the particle approaches zero \cite{Tzarouchis+etal2016}.
In fact, it turns out that the dipole resonance of a small sphere has a very subtle limiting behavior as the electrical size approaches zero,
and as \eg in \cite{Tzarouchis+etal2016} Pad\'{e} approximants are used to reveal new scattering aspects of small spherical particles.
In this paper, an asymptotic analysis based on the Mie theory is employed to study the limiting behavior of the quasistatic optimal resonance \cite{Nordebo+etal2017a},
as the electrical size of the sphere as well as the external losses tend to zero. The limitation of the quasistatic theory is then finally assessed by providing explicit
asymptotic formulas for the validity of the quasistatic model of the optimal resonance.  
We explicitly find the validity region of the quasistatic model, which is determined by the scattering loss factor.

The rest of the paper is organized as follows. In Section \ref{sect:paradox} is given a description of
the quasistatic optimal plasmonic resonance of an ellipsoid embedded in a lossy background medium. 
In Section \ref{sect:Miesolution} we develop a detailed full electrodynamic analysis with explicit asymptotic results
concerning the special case with a sphere.
Numerical examples are given in Section \ref{sect:numexamples} and the vector spherical 
waves are defined in Appendix \ref{sect:spherical}.

\section{The quasistatic optimal plasmonic resonance of an ellipsoid in a lossy background medium}\label{sect:paradox}
\subsection{Notation and conventions}
The following notations and conventions are used in this paper.
Classical electrodynamics is considered where the electric and magnetic field intensities $\bm{E}$ and $\bm{H}$
are given in SI-units \cite{Jackson1999}.  
The time convention for time harmonic fields (phasors) is given by $\eu^{-\iu\omega t}$
where $\omega$ is the angular frequency and $t$ the time. 
Consequently,  the relative permittivity $\epsilon$ of a passive isotropic dielectric material has positive imaginary part.
Let $\mu_0$, $\epsilon_0$, $\eta_0$ and $\mrm{c}_0$ denote the permeability, the permittivity, the wave impedance and
the speed of light in vacuum, respectively, and where $\eta_0=\sqrt{\mu_0/\epsilon_0}$ and $\mrm{c}_0=1/\sqrt{\mu_0\epsilon_0}$.
The wavenumber of vacuum is given by $k_0=\omega\sqrt{\mu_0\epsilon_0}$.
The wavenumber of a homogeneous and isotropic medium with relative permeability $\mu$ and permittivity $\epsilon$ is given by $k=k_0\sqrt{\mu\epsilon}$
and the wavelength $\lambda$ is defined by $k\lambda=2\pi$.
The wave impedance of the same medium is given by $\eta_0\eta$ where $\eta=\sqrt{\mu/\epsilon}$ is the relative wave impedance.
In the following, we will consider only non-magnetic, homogeneous and isotropic dielectric or conducting materials, and hence $\mu=1$ from now on. 
The spherical coordinates are denoted by $(r,\theta,\phi)$, the corresponding unit vectors $(\hat{\bm{r}},\hat{\bm{\theta}},\hat{\bm{\phi}})$,
and the radius vector $\bm{r}=r\hat{\bm{r}}$.
Finally, the real and the imaginary parts, and the complex conjugate of a complex number $\zeta$ are denoted 
$\Re\left\{\zeta\right\}$, $\Im\left\{\zeta\right\}$ and $\zeta^*$, respectively.

\subsection{Optimization under the quasistatic approximation}
The maximal absorption of a small dielectric ellipsoid under the quasistatic approximation can readily be calculated as follows, see also \cite{Nordebo+etal2017a,Dalarsson+etal2017a}.
Consider a small, homogeneous and isotropic dielectric ellipsoid with relative permittivity $\epsilon$ which is embedded in 
a lossy dielectric background medium with relative permittivity $\epsilon_\mrm{b}$. In the quasistatic approximation, 
the polarizability of the ellipsoid with a uniform excitation $\bm{E}_\mrm{i}=E_0\hat{\bm{e}}$ along one of its axes is given by the expression
\begin{equation}\label{eq:alphaellipsoiddef}
\alpha=V\frac{\epsilon-\epsilon_\mrm{b}}{\epsilon_\mrm{b}+L(\epsilon-\epsilon_\mrm{b})},
\end{equation}
where $0<L<1$ is the corresponding depolarizing factor and $V$ the volume of the ellipsoid, \cf \eg \cite{Bohren+Huffman1983}.
Note that $L<1/3$, $L=1/3$ and $L>1/3$ for a prolate spheroid, the sphere and an oblate spheroid, respectively.
The dipole moment of the small ellipsoid is given by $\bm{p}=\epsilon_0\epsilon_\mrm{b}\alpha\bm{E}_\mrm{i}=\int_\Omega\epsilon_0(\epsilon-\epsilon_\mrm{b})\bm{E}\mrm{d}v$
where $\Omega$ denotes the ellipsoidal domain, and since the resulting internal field $\bm{E}$ of the ellipsoid is a constant vector parallel to 
$\bm{E}_\mrm{i}$ \cite{Bohren+Huffman1983}, it follows readily that
\begin{equation}
\bm{E}=\frac{\epsilon_\mrm{b}\alpha}{V(\epsilon-\epsilon_\mrm{b})}\bm{E}_\mrm{i}=\frac{\epsilon_\mrm{b}}{L\left(\epsilon+\epsilon_\mrm{b}\frac{(1-L)}{L} \right)}\bm{E}_\mrm{i}.
\end{equation}
The power absorbed in the ellipsoid can now be calculated from Poynting's theorem as
\begin{multline}\label{eq:Pabsellipsoid1}
P_\mrm{abs}=\frac{\omega\epsilon_0}{2}\Im\{\epsilon\}\int_\Omega\left|\bm{E} \right|^2\mrm{d}v= \\
\frac{\omega\epsilon_0}{2}\frac{\left|\epsilon_\mrm{b}\right|^2}{L^2}\frac{\Im\{\epsilon\}}{\left|\epsilon+\epsilon_\mrm{b}\frac{(1-L)}{L} \right|^2}\left| E_0\right|^2V,
\end{multline}
or 
\begin{equation}\label{eq:Pabsellipsoid2}
P_\mrm{abs}=
\frac{\omega\epsilon_0}{2}\frac{\left|\epsilon_\mrm{b}\right|^2\Im\{\epsilon\}}{\left|\epsilon_\mrm{b}+L(\epsilon-\epsilon_\mrm{b})\right|^2}\left| E_0\right|^2V.
\end{equation}
As \eg, for the sphere ($L=1/3$), the absorption cross section $C_\mrm{abs}$ is obtained by normalizing with the power intensity 
of the incoming plane wave, yielding
\begin{equation}\label{eq:Cabssphere1}
C_\mrm{abs}=\frac{P_\mrm{abs}}{I_\mrm{i}}
=12\pi k_0 a^3\frac{\left|\epsilon_\mrm{b}\right|^2}{\Re\{\sqrt{\epsilon_\mrm{b}}\}}\frac{\Im\{\epsilon\}}{\left|\epsilon+2\epsilon_\mrm{b} \right|^2},
\end{equation}
where $I_\mrm{i}=\frac{1}{2}\Re\{E_0H_0^*\}=\left| E_0\right|^2\Re\{\sqrt{\epsilon_\mrm{b}}\}/2\eta_0$ and where $H_0=E_0/\eta$.

When the background medium as well as the shape of the ellipsoid are fixed, the expression \eqref{eq:Pabsellipsoid1} can be maximized as follows.
Let $\epsilon_\mrm{opt}$ denote a fixed complex-valued constant with $\Im\{\epsilon_\mrm{opt}\}>0$ and consider the real-valued function 
\begin{equation}\label{eq:fofepsilondef}
f(\epsilon)=\frac{\Im\{\epsilon\}}{\left|\epsilon-\epsilon_\mrm{opt}^* \right|^2},
\end{equation}
where $\epsilon$ is a complex-valued variable with $\Im\{\epsilon\}>0$.
It can be shown that the function $f(\epsilon)$ has a local maximum at $\epsilon_\mrm{opt}$ \cf \cite[Sect.~2.5, Eqs.~(15) through (17)]{Nordebo+etal2017a}.
Hence, for the ellipsoid the maximizer of $P_\mrm{abs}$ is given by
\begin{equation}\label{eq:epsilonodef}
\epsilon_\mrm{opt}=-\epsilon_\mrm{b}^*\frac{(1-L)}{L},
\end{equation}
and in particular for the sphere $\epsilon_\mrm{opt}=-2\epsilon_\mrm{b}^*$, 
\cf \cite{Nordebo+etal2017a,Dalarsson+etal2017a}.
The corresponding maximal absorption is given by
\begin{equation}\label{eq:Pabsellipsoid1o}
P_\mrm{abs}^\mrm{qs,opt}(\epsilon_\mrm{b},L)=\frac{\omega\epsilon_0}{2}\frac{1}{4L(1-L)}\frac{\left|\epsilon_\mrm{b}\right|^2}{\Im\{\epsilon_\mrm{b}\}}\left| E_0\right|^2V,
\end{equation}
and for the sphere we obtain the optimal absorption cross section
\begin{equation}\label{eq:Cabssphere1o}
C_\mrm{abs}^\mrm{qs,opt}
=\frac{3\pi}{2}k_0 a^3\frac{\left|\epsilon_\mrm{b}\right|^2}{\Re\{\sqrt{\epsilon_\mrm{b}}\}}\frac{1}{\Im\{\epsilon_\mrm{b}\}}.
\end{equation}
The solution \eqref{eq:epsilonodef} is referred to as an optimal conjugate match and can be
interpreted in terms of an optimal plasmonic resonance for the ellipsoid \cite{Nordebo+etal2017a,Dalarsson+etal2017a}.
Note that \eqref{eq:Pabsellipsoid1o} is unbounded in the cases when the exterior domain becomes lossless and $\Im\{\epsilon_\mrm{b}\}\rightarrow 0$,
as well as when the ellipsoid collapses and $L\rightarrow 0$ or $L\rightarrow 1$.

When both media parameters $\epsilon$ and $\epsilon_\mrm{b}$ are fixed, the expression \eqref{eq:Pabsellipsoid2} can be maximized
with respect to the shape parameter $L$. By straightforward differentiation it is found that the maximizing shape parameter $L_\mrm{opt}$ is given by
\begin{equation}\label{eq:Lodef}
L_\mrm{opt}=\Re\left\{\frac{\epsilon_\mrm{b}}{\epsilon_\mrm{b}-\epsilon} \right\},
\end{equation}
and the corresponding maximal absorption is given by
\begin{equation}\label{eq:Pabsellipsoid2o}
P_\mrm{abs}^\mrm{qs,opt}(\epsilon_\mrm{b},\epsilon)
=\frac{\omega\epsilon_0}{2}\frac{\left|\epsilon_\mrm{b}\right|^2\Im\{\epsilon\}\left|\epsilon-\epsilon_\mrm{b}\right|^2}{\left(\Im\{\epsilon_\mrm{b}^*\epsilon\} \right)^2}
\left| E_0\right|^2V.
\end{equation}
In the limiting case when the exterior region becomes lossless and $\Im\{\epsilon_\mrm{b}\}\rightarrow 0$, the expression \eqref{eq:Pabsellipsoid2o} simplifies to
\begin{equation}\label{eq:Pabsellipsoid3o}
P_\mrm{abs}^\mrm{qs,opt}(\epsilon_\mrm{b},\epsilon)
=\frac{\omega\epsilon_0}{2}\frac{\left|\epsilon-\epsilon_\mrm{b}\right|^2}{\Im\{\epsilon\}}
\left| E_0\right|^2V,
\end{equation}
and which agrees with the upper bound given in \cite[Eq.~(32b) on p.~3345 and Eq.~(41) on p.~3349]{Miller+etal2016} under the same quasistatic assumption (incident field is uniform).

What is important to note at this point is the unboundedness of the quasistatic maximal absorption \eqref{eq:Pabsellipsoid1o} as the external loss factor
$\Im\{\epsilon_\mrm{b}\}$ tends to zero. Obviously, this is in contradiction to the well known bound on the 
absorption of an arbitrary electric dipole scatterer in a lossless background medium given by
\begin{equation}\label{eq:Cabsdip}
C_\mrm{abs}^\mrm{dip}=\frac{3\pi}{2}\frac{1}{k_\mrm{b}^2},
\end{equation}
where $k_\mrm{b}=k_0\sqrt{\epsilon_\mrm{b}}$, \cf \cite[Eq.~(16) on p.~937]{Tretyakov2014}. As we will see later, this apparent contradiction is due simply
to the fact that the quasistatic approximation does not take the scattering loss into account. In particular, in the next section we will see
that the fully dynamical model for the normalized absorption cross section area of a small dielectric spherical dipole has a very subtle limiting behavior, 
and is in fact unbounded when $\epsilon=-2\epsilon_\mrm{b}^*$ and both the electrical size as well as the exterior losses tend to zero at the same time. 
To this end, it is also interesting to observe that the limit of \eqref{eq:Pabsellipsoid2o} as $(\Im\{\epsilon\},\Im\{\epsilon_\mrm{b}\})\rightarrow (0,0)$ will depend on how
this limit is taken. In particular, \eqref{eq:Pabsellipsoid3o} is obtained for fixed $\epsilon$ ($\Im\{\epsilon\}>0$) as $\Im\{\epsilon_\mrm{b}\}\rightarrow 0$ and \eqref{eq:Pabsellipsoid3o} is
then unbounded as $\Im\{\epsilon\}$ approaches zero. On the other hand, \eqref{eq:Pabsellipsoid2o} approaches zero for fixed  $\Im\{\epsilon_\mrm{b}\}>0$ as  $\Im\{\epsilon\}\rightarrow 0$ 
(and $\Re\{\epsilon\}\neq 0$).

The conclusion of this discussion is that one can not optimize (with respect to $\epsilon$) the absorption of an ellipsoid under the quasistatic assumption
when the exterior domain is lossless. The natural question that arises is then under which circumstances this optimization model is valid for a lossy exterior domain.
This is the topic of the next section, where we restrict the analysis to the absorption of a dielectric sphere in a lossy medium.

\section{The absorption of a small dielectric sphere in a lossy background medium}\label{sect:Miesolution}
\subsection{Electrodynamic solution}
The complete electrodynamic solution for the internal absorption of a small dielectric sphere in a lossy background medium is analyzed below.
The definition of the spherical vector waves, the spherical Bessel and Hankel functions and the related Lommel integrals are given in Appendix \ref{sect:spherical}, see also  \cite{Nordebo+etal2017a}.

Consider the scattering of the electromagnetic field due to a homogeneous dielectric sphere of radius $a$, complex-valued permittivity $\epsilon$, 
and wavenumber $k=k_0\sqrt{\epsilon}$.
The medium surrounding the sphere is characterized by the permittivity $\epsilon_\mrm{b}$ and the wavenumber $k_\mrm{b}=k_0\sqrt{\epsilon_\mrm{b}}$.
The incident and the scattered fields for $r>a$ are expressed as in \eqref{eq:Esphdef} with multipole coefficients $a_{\tau ml}$ and $b_{\tau ml}$, respectively,
and the interior field is similarly expressed using regular spherical vector waves for $r<a$ with multipole coefficients $a_{\tau ml}^\mrm{int}$.
By matching the tangential electric and magnetic fields at the boundary of radius $a$, it can be shown that
\begin{eqnarray}\label{eq:btaumldef}
b_{\tau ml}=t_{\tau l}a_{\tau ml}, \\
a_{\tau ml}^\mrm{int}=r_{\tau l}a_{\tau ml},\label{eq:ataumldef}
\end{eqnarray}
for $\tau=1,2$, $l=1,2,\ldots$, and $m=-l,\ldots,l$, and where $t_{\tau l}$ and $r_{\tau l}$ are transition matrices for scattering and absorption, 
respectively, see \eg \cite[Eqs.~(4.52) and (4.53) on p.~100]{Bohren+Huffman1983}.
In particular, for the internal fields of the sphere the corresponding electric multipole coefficients are given by
\begin{multline} 
r_{2l} = \\
 \displaystyle  \frac{ (\mrm{j}_l(k_\mrm{b}a) (k_\mrm{b}a\mrm{h}_l^{(1)}(k_\mrm{b}a))^\prime-\mrm{h}_l^{(1)}(k_\mrm{b}a) (k_\mrm{b}a\mrm{j}_l(k_\mrm{b}a))^\prime)\sqrt{\epsilon_\mrm{b}\epsilon} }
   { -\mrm{h}_l^{(1)}(k_\mrm{b}a) (ka\mrm{j}_l(ka))^\prime\epsilon_\mrm{b} + \mrm{j}_l(ka) (k_\mrm{b}a\mrm{h}_l^{(1)}(k_\mrm{b}a))^\prime\epsilon}.  \label{eq:r2l}
\end{multline}

It can be shown that the multipole expansion coefficients for a plane
wave $\bm{E}_\mrm{i}(\bm r)=\bm{E}_0\eu^{i k_\mrm{b}\hat{\bm{k}}\cdot\bm{r}}$ are given by
\begin{equation}
a_{\tau ml}=4\pi\iu^{l-\tau+1}\bm{E}_0\cdot\bm{A}_{\tau ml}^*(\hat{\bm{k}}),
\end{equation}
for $\tau=1,2$, $l=1,2,\ldots$, and $m=-l,\ldots,l$, and where the vector spherical harmonics $\bm{A}_{\tau ml}(\hat{\bm{k}})$
are defined as in Appendix \ref{sect:sphericaldef}, \cf \eg \cite[Eq.~(7.27) on p.~374]{Kristensson2016}. Without loss of generality, 
for a homogeneous and isotropic sphere we may assume that $\hat{\bm{k}}=\hat{\bm{z}}$, and hence from \cite[Eq.~(7.29) on p.~375]{Kristensson2016} it follows that 
\begin{equation}\label{eq:sumoverm}
\sum_{m=-l}^l \left| a_{\tau ml} \right|^2=2\pi (2l+1)\left| \bm{E}_0\right|^2.
\end{equation}

Based on Poynting's theorem and the orthogonality of the spherical vector waves
the absorbed power can now be calculated as
\begin{multline}\label{eq:Pabssphexpdef}
P_\mrm{abs}=\frac{1}{2}\omega\epsilon_0\Im\{\epsilon\}\int_{V_{a}} \left| \bm{E} \right|^2\mrm{d}v  \\
 =\pi\left| \bm{E}_0\right|^2\omega\epsilon_0\Im\{\epsilon\}\sum_{l=1}^{\infty}\sum_{\tau=1}^2  (2l+1) W_{\tau l}(k,a)\left| r_{\tau l} \right|^2, 
\end{multline}
where $V_{a}$ denotes the spherical volume of radius $a$ and where 
$W_{\tau l}(k,a)=\int_{V_{a}}\left|{\bf v}_{\tau ml}(k\bm{r})\right|^2\mrm{d} v$ is the volume integral of the regular spherical vector waves, 
\cf \eqref{eq:vorthogonal2} and \eqref{eq:Wtauldef}. 
In \eqref{eq:Pabssphexpdef}, the relations \eqref{eq:ataumldef} and \eqref{eq:sumoverm} have also been employed.

The absorption cross section ($C_\mrm{abs}=P_\mrm{abs}/I_\mrm{i}$)
based solely on the dominating Transverse Magnetic (TM) dipole fields ($\tau=2$ and $l=1$) can now be expressed as
\begin{equation}\label{eq:CabssphTMdip}
C_\mrm{abs}^\mrm{dyn}=6\pi k_0 \frac{\Im\{\epsilon\}}{\Re\{\sqrt{\epsilon_\mrm{b}}\}}W_{21}\left| r_{21} \right|^2,
\end{equation}
where 
\begin{multline}
r_{21} =   \\
 \displaystyle  \frac{ (\mrm{j}_1(k_\mrm{b}a) (k_\mrm{b}a\mrm{h}_1^{(1)}(k_\mrm{b}a))^\prime-\mrm{h}_1^{(1)}(k_\mrm{b}a) (k_\mrm{b}a\mrm{j}_1(k_\mrm{b}a))^\prime)\sqrt{\epsilon_\mrm{b}\epsilon} }
   { -\mrm{h}_1^{(1)}(k_\mrm{b}a) (ka\mrm{j}_1(ka))^\prime\epsilon_\mrm{b} + \mrm{j}_1(ka) (k_\mrm{b}a\mrm{h}_1^{(1)}(k_\mrm{b}a))^\prime\epsilon}, \label{eq:r21}
 \end{multline}
 is given by \eqref{eq:r2l}, and
\begin{equation}\label{eq:W21expr}
W_{21}=\frac{a^2}{3}\frac{\Im\left\{k\left( 2\mrm{j}_1(ka)\mrm{j}_0^*(ka)+ \mrm{j}_3(ka)\mrm{j}_2^*(ka) \right)\right\}}{\Im\{k^2\}},
\end{equation}
is obtained from \eqref{eq:W1ldef} and \eqref{eq:W2ldef}.

\subsection{Asymptotic analysis}\label{sect:asymptoticanalys}
To analyze the asymptotic properties of the dynamic solution $C_\mrm{abs}^\mrm{dyn}$, the expression \eqref{eq:CabssphTMdip}
is first rewritten in the following area normalized form
\begin{equation}\label{eq:CabssphTMdipnorm}
Q_\mrm{abs}^\mrm{dyn}=\frac{C_\mrm{abs}^\mrm{dyn}}{\pi a^2}=6 k_0a \frac{\Im\{\epsilon\}}{\Re\{\sqrt{\epsilon_\mrm{b}}\}}\frac{W_{21}}{a^3}\left| r_{21} \right|^2.
\end{equation}
An asymptotic analysis can now be carried out based on the asymptotic expansions of the spherical Bessel and Hankel functions for small arguments, see \eg \cite{Olver+etal2010}.
Furthermore, with $\epsilon=\epsilon^\prime+\iu\epsilon^{\prime\prime}$ and 
$\epsilon_\mrm{b}=\epsilon_\mrm{b}^\prime+\iu\epsilon_\mrm{b}^{\prime\prime}$ and small loss factors $\epsilon^{\prime\prime}$
and $\epsilon_\mrm{b}^{\prime\prime}$, it is readily seen that
\begin{equation}
\left\{\begin{array}{l}
ka=k_0a\sqrt{\epsilon}=k_0a\left(\sqrt{\epsilon^{\prime}}+{\cal O}\{\epsilon^{\prime\prime}\} \right), \vspace{0.2cm} \\
k_\mrm{b}a=k_0a\sqrt{\epsilon_\mrm{b}}=k_0a\left(\sqrt{\epsilon_\mrm{b}^{\prime}}+{\cal O}\{\epsilon_\mrm{b}^{\prime\prime}\} \right),
\end{array}\right.
\end{equation}
where ${\cal O}\{\cdot\}$ denotes the big ordo \cite[p.~4]{Olver1997},
and where it is assumed that $\epsilon^\prime\neq 0$ and $\epsilon_\mrm{b}^\prime\neq 0$ are fixed. 
From this observation, it is also noted that 
for integer $m$
\begin{equation}
\left\{\begin{array}{l}
{\cal O}\{(ka)^m\}={\cal O}\{(k_0a)^m\}, \vspace{0.2cm} \\
 {\cal O}\{(k_\mrm{b}a)^m\}={\cal O}\{(k_0a)^m\},
\end{array}\right.
\end{equation}
independent of the (small) values of $\epsilon^{\prime\prime}$ and $\epsilon_\mrm{b}^{\prime\prime}$.

Based on the asymptotic tools mentioned above, the different factors of \eqref{eq:CabssphTMdipnorm} can now be analyzed in detail.
We start by studying the normalized factor
\begin{equation}\label{eq:W21normexpr}
\frac{W_{21}}{a^3}=\frac{1}{3k_0a}\frac{\Im\left\{ F(k_0a,\epsilon)\right\}}{\Im\{\epsilon\}},
\end{equation}
where
\begin{multline}
F(k_0a,\epsilon) 
=\sqrt{\epsilon}\left( 2\mrm{j}_1(k_0a\sqrt{\epsilon})\mrm{j}_0^*(k_0a\sqrt{\epsilon}) \right.  \\
 + \left. \mrm{j}_3(k_0a\sqrt{\epsilon})\mrm{j}_2^*(k_0a\sqrt{\epsilon}) \right)
\end{multline}
is given by \eqref{eq:W21expr}.
Note that the spherical Bessel functions $\mrm{j}_n(x)$ of order $n$ are analytic functions with a Taylor series expansion starting at the polynomial order $n$.
The functions $\mrm{j}_n(x)$ are furthermore odd for odd orders and even for even orders. Hence, it is readily seen that the function
$F(k_0a,\epsilon)$ has a convergent power series expansion in terms of $\epsilon$ and $\epsilon^*$ with real valued coefficients, and that
\begin{eqnarray}
F(k_0a,\epsilon) =k_0a\frac{2}{3}\epsilon+{\cal O}\{\left(k_0a\right)^3\}{\cal O}\{\epsilon\},
\end{eqnarray}
and 
\begin{eqnarray}
\Im\{F(k_0a,\epsilon)\} =k_0a\frac{2}{3}\epsilon^{\prime\prime}+{\cal O}\{\left(k_0a\right)^3\}{\cal O}\{\epsilon^{\prime\prime}\}.
\end{eqnarray}
This means that the expression in \eqref{eq:W21normexpr} is well behaved when $k_0a$ as well as $\epsilon^{\prime\prime}$ approaches zero,
and that
\begin{equation}\label{eq:W21normas}
\frac{W_{21}}{a^3}=\frac{2}{9}+{\cal O}\{(k_0a)^2\},
\end{equation}
independent of $\epsilon^{\prime\prime}$. This observation can also be derived directly from the definition \eqref{eq:W2ldef} and the regularity
of the spherical Bessel function $\mrm{j}_1(x)$.

A detailed asymptotic study of \eqref{eq:r21} shows that
\begin{equation}\label{eq:r21as}
r_{21}=\frac{3\epsilon_\mrm{b}+{\cal O}\{(k_0a)^2\}}{\epsilon+2\epsilon_\mrm{b}+(k_0a)^2C+(k_0a)^3D+ {\cal O}\{(k_0a)^4\}},
\end{equation}
where 
\begin{equation}\label{eq:CDdef}
\left\{\begin{array}{l}
C=(\epsilon_\mrm{b}-\epsilon)(\epsilon_\mrm{b}+\epsilon/10), \vspace{0.2cm} \\
D=\iu\frac{2}{3}\epsilon_\mrm{b}\sqrt{\epsilon_\mrm{b}}(\epsilon_\mrm{b}-\epsilon). 
\end{array}\right.
\end{equation}
It can safely be assumed that $C\neq 0$ and $D\neq 0$ since
$\epsilon_\mrm{b}=\epsilon$ is not an interesting case and $\epsilon_\mrm{b}+\epsilon/10\neq 0$ for lossy and passive media.
For fixed $\epsilon$ and $\epsilon_\mrm{b}$ with $\epsilon+2\epsilon_\mrm{b}\neq 0$, and by using \eqref{eq:W21normas} and \eqref{eq:r21as}, 
it is now concluded that \eqref{eq:CabssphTMdip} is given by
\begin{equation}\label{eq:CabssphTMdipas}
C_\mrm{abs}^\mrm{dyn}=12\pi k_0 a^3\frac{\left|\epsilon_\mrm{b}\right|^2}{\Re\{\sqrt{\epsilon_\mrm{b}}\}}
\frac{\Im\{\epsilon\}}{\left|\epsilon+2\epsilon_\mrm{b} \right|^2}+{\cal O}\{(k_0a)^2\},
\end{equation}
which is in agreement with the quasistatic approximation given in \eqref{eq:Cabssphere1}. 

However, when $\epsilon^{\prime}$ and $\epsilon_\mrm{b}^{\prime}$ are fixed at $\epsilon^{\prime}=-2\epsilon_\mrm{b}^{\prime}$,
and at the same time $k_0a$ as well as $\epsilon^{\prime\prime}$ and $\epsilon_\mrm{b}^{\prime\prime}$ can be chosen freely, 
the normalized  absorption cross section area \eqref{eq:CabssphTMdipnorm} turns out to be unbounded. To study this behavior in detail, it is noted that to the lowest order in
$(k_0a,\epsilon^{\prime\prime},\epsilon_\mrm{b}^{\prime\prime})$, the expression \eqref{eq:CabssphTMdipnorm} can be approximated by
\begin{equation}\label{eq:CabssphTMdipnormapprox1}
Q_\mrm{abs}^\mrm{dyn}\sim 12 k_0a \frac{\left|\epsilon_\mrm{b}\right|^2}{\Re\{\sqrt{\epsilon_\mrm{b}}\}}
\frac{\Im\{\epsilon\}}{\left|\epsilon+2\epsilon_\mrm{b}+(k_0a)^2C_0\right|^2},
\end{equation}
where $C_0=(\epsilon_\mrm{b}^\prime-\epsilon^\prime)(\epsilon_\mrm{b}^\prime+\epsilon^\prime/10)$ and the symbol $\sim$ indicates
an asymptotic approximation in the sense of \cite[p.~4]{Olver1997}.
Obviously, for any small but finite value of $k_0a\neq 0$ the real part of the expression under the absolute-value sign can be brought to zero, 
and we see that in this approximation the absorption cross section is unbounded at $\epsilon^{\prime\prime}$ and 
$\epsilon_b^{\prime\prime}$ tending to zero. We again confirm that the second-order correction is not enough to reveal the dynamic absorption bound.
Interestingly, the absorption cross section can be unbounded even without tuning to the dynamically-corrected resonant value of the permittivity. 
In particular, by considering the special case of interest, \ie the quasistatic optimal conjugate match where 
$\epsilon=-2\epsilon_\mrm{b}^*=-2\epsilon_\mrm{b}^\prime+\iu 2\epsilon_\mrm{b}^{\prime\prime}$
and $\epsilon+2\epsilon_\mrm{b}=\iu 4 \epsilon_\mrm{b}^{\prime\prime}$, it follows that
\begin{equation}\label{eq:CabssphTMdipnormapprox2}
Q_\mrm{abs}^\mrm{dyn}\sim \frac{3}{2} \frac{\left|\epsilon_\mrm{b}\right|^2}{\Re\{\sqrt{\epsilon_\mrm{b}}\}}
\frac{k_0a\epsilon_\mrm{b}^{\prime\prime}}{{\epsilon_\mrm{b}^{\prime\prime}}^2+(k_0a)^4C_0^2/16},
\end{equation}
where $C_0=12{\epsilon_\mrm{b}^\prime}^2/5$ and where it is assumed that $\epsilon_\mrm{b}^\prime\neq 0$. Note that the assumption
$\epsilon_\mrm{b}^{\prime\prime} \gg (k_0a)^2C_0/4$ will restore the expression for the maximal absorption given by \eqref{eq:Cabssphere1o}.
Hence, it is observed that the quasistatic theory \eqref{eq:Cabssphere1o} is valid provided that $k_0a$ is sufficiently small
at the same time as the background loss factor $\epsilon_\mrm{b}^{\prime\prime}$ is sufficiently large.

The factor governing the convergence of \eqref{eq:CabssphTMdipnormapprox2} and \eqref{eq:CabssphTMdipnorm} is given by
\begin{equation}\label{eq:factorgovconv1}
F_1=\frac{k_0a\epsilon_\mrm{b}^{\prime\prime}}{{\epsilon_\mrm{b}^{\prime\prime}}^2+(k_0a)^4C_0^2/16}.
\end{equation}
By choosing 
\begin{equation}\label{eq:k0aasymptotics}
k_0a=A{\epsilon_\mrm{b}^{\prime\prime}}^\alpha,
\end{equation}
where $A$ is an arbitrary constant and $\alpha>0$, this factor becomes
\begin{equation}\label{eq:factorgovconv2}
F_2=\frac{A{\epsilon_\mrm{b}^{\prime\prime}}^{\alpha+1}}{{\epsilon_\mrm{b}^{\prime\prime}}^2+A^4{\epsilon_\mrm{b}^{\prime\prime}}^{4\alpha}C_0^2/16}.
\end{equation}
A detailed study of the expression \eqref{eq:factorgovconv2} for small $\epsilon_\mrm{b}^{\prime\prime}>0$ 
reveals the condition for convergence of \eqref{eq:CabssphTMdipnorm}, which can be summarized as
\begin{equation}\label{eq:convdiv}
\left\{\begin{array}{ll}
\mrm{Convergence} & 0<\alpha<\frac{1}{3}, \vspace{0.2cm} \\
\mrm{Divergence} & \frac{1}{3}\leq \alpha\leq 1, \vspace{0.2cm} \\
\mrm{Convergence} & \alpha>1, 
\end{array}\right.
\end{equation}
where $k_0a=A{ \epsilon_\mrm{b}^{\prime\prime}}^\alpha$, $\epsilon_\mrm{b}^\prime$ is fixed and $\epsilon=-2\epsilon_\mrm{b}^*$ (the quasistatic optimal conjugate match).

For a fixed background loss parameter $\epsilon_\mrm{b}^{\prime\prime}>0$, the factor \eqref{eq:factorgovconv1} can furthermore be maximized
with respect to the electrical size $k_0a$, yielding
\begin{equation}\label{eq:k0aoptasymptotics}
k_0a=\frac{2}{3^{1/4}C_0^{1/2}}{\epsilon_\mrm{b}^{\prime\prime}}^{1/2}.
\end{equation}
For fixed $\epsilon_\mrm{b}^{\prime\prime}$, the relation \eqref{eq:k0aoptasymptotics} expresses a stationary point
for \eqref{eq:CabssphTMdipnormapprox2} regarded as a function of $k_0a$, and hence an indicator of the domain of validity of 
the quasistatic approximation \eqref{eq:Cabssphere1o}.

In conclusion, it has been shown that the  normalized absorption cross section area
\eqref{eq:CabssphTMdipnorm} has a subtle limiting behavior and is in fact unbounded as 
$(k_0a,\epsilon^{\prime\prime},\epsilon_\mrm{b}^{\prime\prime})\rightarrow (0,0,0)$ for fixed $\epsilon^{\prime}$ and $\epsilon_\mrm{b}^{\prime}$.
It is also shown that if $k_0a$ is sufficiently small, the optimal conjugate match $\epsilon=-2\epsilon_\mrm{b}^*$ defined in \eqref{eq:epsilonodef} yields
an accurate quasistatic approximation \eqref{eq:Cabssphere1o} provided that 
\begin{equation}\label{eq:k0aoptasymptotics2}
\epsilon_\mrm{b}^{\prime\prime}>\frac{3\sqrt{3}}{5}\left(k_0a\right)^2{\epsilon_\mrm{b}^\prime}^2,
\end{equation}
where \eqref{eq:k0aoptasymptotics} and $C_0=12{\epsilon_\mrm{b}^\prime}^2/5$ have been used.
It should be noted that a direct maximization of \eqref{eq:CabssphTMdipnorm} with respect to $\epsilon$ would be possible by numerical optimization techniques,
but is not necessary if the quasistatic approximation is valid. The validity of the quasistatic approximation can be assessed by checking
the criteria \eqref{eq:k0aoptasymptotics2} as well as by a direct comparison of \eqref{eq:Cabssphere1} and \eqref{eq:Cabssphere1o} with
the electrodynamic solution \eqref{eq:CabssphTMdipnorm}.

\subsection{Optimal absorption of the small dielectric sphere in a lossy media}\label{sect:optimalabsorptionepsilonsol}
Finally, the asymptotic analysis above with \eqref{eq:r21as} and \eqref{eq:CDdef} makes it possible to analyze the pole structure of
\eqref{eq:CabssphTMdipnorm} in full dynamics. By making the following Ansatz for the pole
$\epsilon_\mrm{p}=a_0+a_1 k_0a +a_2(k_0a)^2+a_3(k_0a)^3$,
and identifying terms up to the third order in the denominator of \eqref{eq:r21as},
it is found that \eqref{eq:CabssphTMdipnorm} can be approximated for small $k_0a$ as
\begin{equation}\label{eq:Qabsdynpoleapprox}
Q_\mrm{abs}^\mrm{dyn}\sim 12 k_0a \frac{\left| \epsilon_\mrm{b} \right|^2}{\Re\{\sqrt{\epsilon_\mrm{b}}\}}  \frac{\Im\{\epsilon\}}{\left|\epsilon-\epsilon_\mrm{p} \right|^2},
\end{equation}
and where the pole is given by 
\begin{equation}\label{eq:epsilonpolespherethirdorder}
\epsilon_\mrm{p}=-2\epsilon_\mrm{b}-\frac{12}{5}\epsilon_\mrm{b}(k_\mrm{b}a)^2-\iu 2 \epsilon_\mrm{b}(k_\mrm{b}a)^3+ {\cal O}\{(k_\mrm{b}a)^4\},
\end{equation}
see also \cite{Tretyakov2014}, and \cite[Eq.~(11) on p.~3]{Tzarouchis+etal2016}.
The expression \eqref{eq:Qabsdynpoleapprox} is of the form \eqref{eq:fofepsilondef} where $\Im\{\epsilon_\mrm{p}\}<0$ for small $\epsilon_\mrm{b}^{\prime\prime}$, 
and hence it can be concluded that the maximal absorption is approximately (asymptotically) achieved at $\epsilon_\mrm{opt}=\epsilon_\mrm{p}^*$, yielding
\begin{equation}\label{eq:epsilonospherethirdorder}
\epsilon_\mrm{opt}=-2\epsilon_\mrm{b}^*-\frac{12}{5}\epsilon_\mrm{b}^*(k_\mrm{b}^*a)^2+\iu 2 \epsilon_\mrm{b}^*(k_\mrm{b}^*a)^3+ {\cal O}\{(k_\mrm{b}^*a)^4\}.
\end{equation}
Note that the result \eqref{eq:epsilonospherethirdorder} generalizes the previous quasistatic result $\epsilon_\mrm{opt}=-2\epsilon_\mrm{b}^*$ given by \eqref{eq:epsilonodef}.
The expression is valid for small $k_0a$ as well as for small loss factors $\epsilon_\mrm{b}^{\prime\prime}$.
In particular, the optimal absorption of the sphere is given by
\begin{equation}\label{eq:Qabsoptasymptotic}
Q_\mrm{abs}^\mrm{dyn,opt}\sim 3k_0a\frac{\left| \epsilon_\mrm{b} \right|^2}{\Re\{\sqrt{\epsilon_\mrm{b}}\}}\frac{1}{\Im\{\epsilon_\mrm{opt}\}},
\end{equation}
and in the limit as the exterior losses vanish, 
we have that $\Im\{\epsilon_\mrm{opt}\}\rightarrow 2{\epsilon_\mrm{b}^\prime}^2\sqrt{\epsilon_\mrm{b}^\prime}(k_0a)^3$ and
\begin{equation}\label{eq:Qabsoptasymptotic2}
\lim_{\epsilon_\mrm{b}^{\prime\prime}\rightarrow 0}Q_\mrm{abs}^\mrm{dyn,opt}=\frac{3}{2}\frac{1}{(k_\mrm{b} a)^2},
\end{equation}
where $k_\mrm{b}=k_0\sqrt{\epsilon_\mrm{b}^\prime}$, and which is in agreement with the upper bound \eqref{eq:Cabsdip} given in \cite{Tretyakov2014}.

Finally, it should be noted that there are major discrepancies in the asymptotic analysis 
of the full dynamic solution performed in this section in comparison to the analysis of the quasistatic
approximation as in \eqref{eq:CabssphTMdipnormapprox1} and \eqref{eq:CabssphTMdipnormapprox2} where $\epsilon^{\prime}$ and $\epsilon_\mrm{b}^{\prime}$ are fixed 
and $\epsilon^{\prime}=-2\epsilon_\mrm{b}^{\prime}$.
In essence, the quasistatic approximation \eqref{eq:epsilonodef} is missing the second-order term giving a shift in the resonance frequency as well as the
third-order term taking the scattering loss into account. 
These factors are negligible only when the external losses are large enough  to make the second-order term redundant as expressed in \eqref{eq:k0aoptasymptotics2},
at the same time as the electrical size of the sphere is small enough to make the scattering loss insignificant. 
Hence, when the external losses are too small in relation to the electrical size of the sphere, 
the quasistatic solution $\epsilon_\mrm{opt}=-2\epsilon_\mrm{b}^*$ is not the correct choice to maximize the absorption and
one should instead use \eqref{eq:epsilonospherethirdorder}.

\section{Numerical examples}\label{sect:numexamples}

\begin{figure}[htb]
\begin{center}
\includegraphics[width=0.48\textwidth]{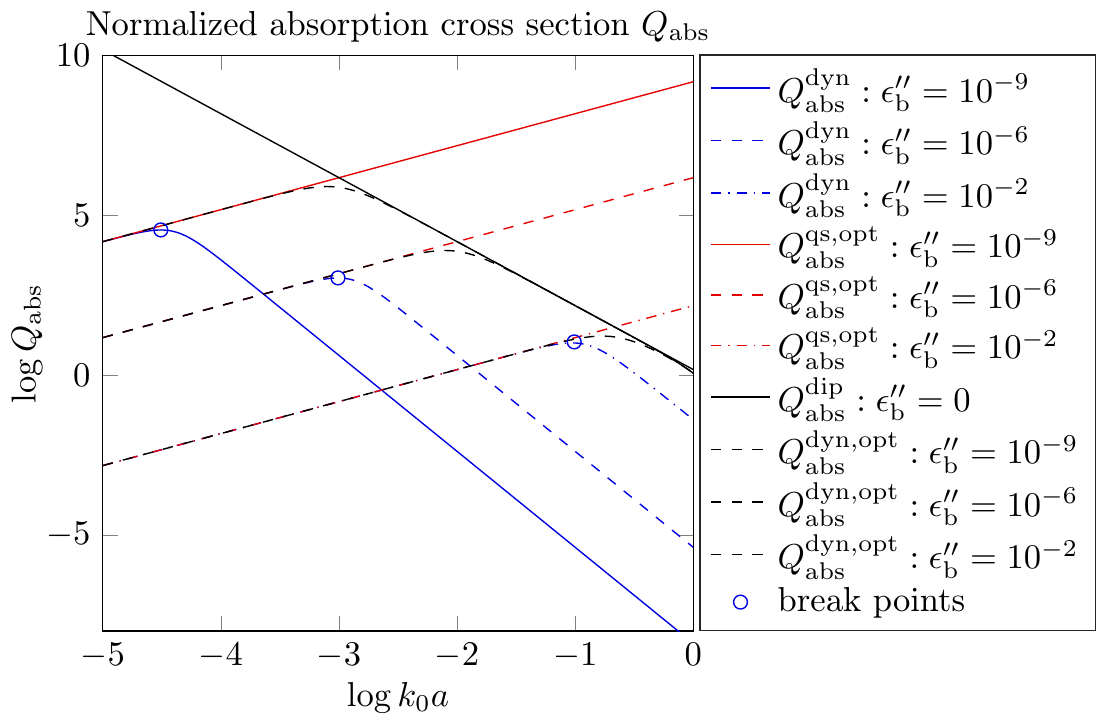}
\end{center}
\vspace{-5mm}
\caption{The quasistatic optimal normalized absorption cross section $Q_\mrm{abs}^\mrm{qs,opt}$ and the corresponding dynamic cross section $Q_\mrm{abs}^\mrm{dyn}$ 
as functions of the electrical size $k_0a$. The break points are given by the asymptotic analysis
\eqref{eq:k0aoptasymptotics}. For comparison, the optimal dynamic $Q_\mrm{abs}^\mrm{dyn,opt}$ is calculated based on \eqref{eq:epsilonospherethirdorder}.
Here, $\epsilon_\mrm{b}^{\prime}=1$.}
\label{fig:matfig3}
\end{figure}

\begin{figure}[htb]
\begin{center}
\includegraphics[width=0.48\textwidth]{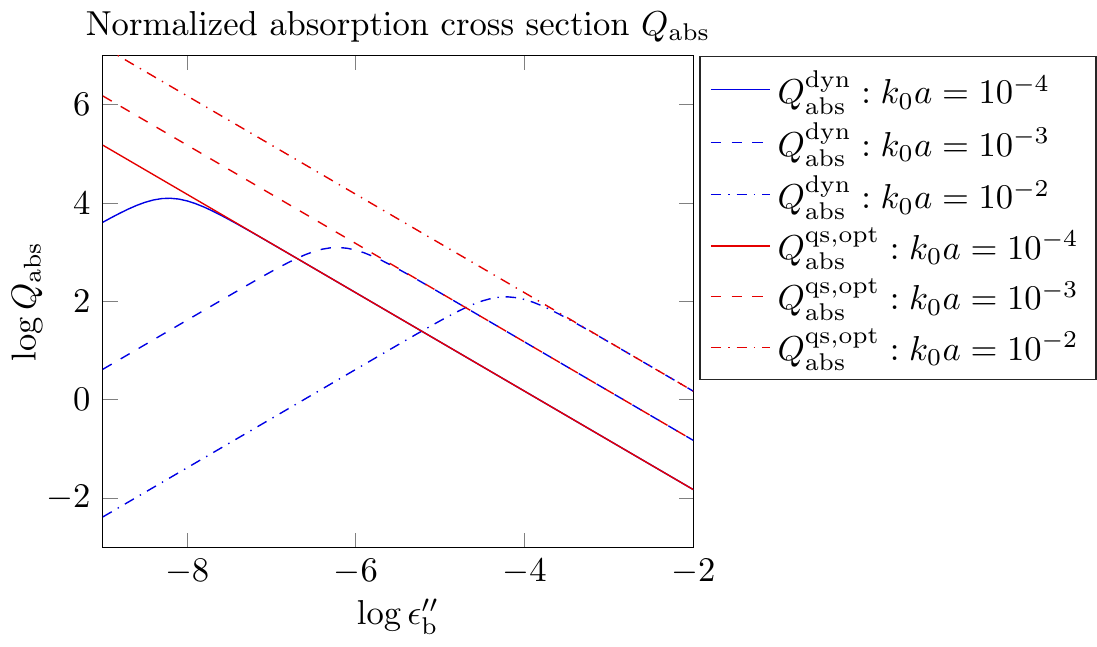}
\end{center}
\vspace{-5mm}
\caption{The quasistatic optimal normalized absorption cross section $Q_\mrm{abs}^\mrm{qs,opt}$ and the corresponding dynamic cross section $Q_\mrm{abs}^\mrm{dyn}$  
as functions of the background loss $\epsilon_\mrm{b}^{\prime\prime}$. Here, $\epsilon_\mrm{b}^{\prime}=1$.}
\label{fig:matfig4}
\end{figure}

\begin{figure}[htb]
\begin{center}
\includegraphics[width=0.48\textwidth]{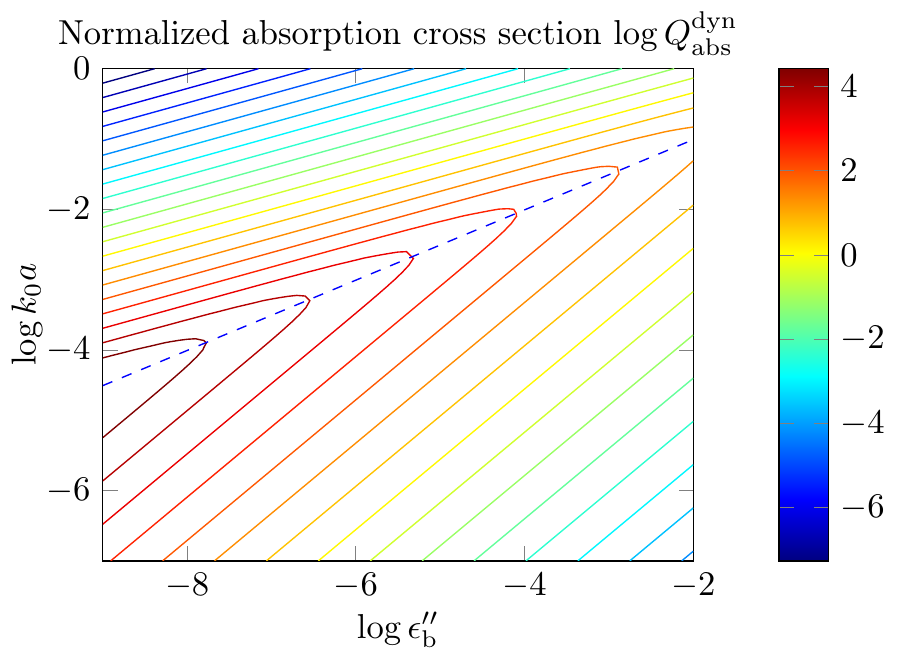}
\end{center}
\vspace{-5mm}
\caption{The normalized dynamic absorption cross section $Q_\mrm{abs}^\mrm{dyn}$ as a function of the electrical size $k_0a$ 
and the background loss $\epsilon_\mrm{b}^{\prime\prime}$, and where $\epsilon_\mrm{b}^{\prime}=1$. 
The blue dashed line shows the stationary (break) points given by the asymptotic analysis
\eqref{eq:k0aoptasymptotics} indicating the region of validity \eqref{eq:k0aoptasymptotics2} of the quasistatic approximation.
Note also the asymptotes for divergence given by $\log k_0a=\log A+\alpha\log { \epsilon_\mrm{b}^{\prime\prime}}$ with $\alpha=1/3$ and $\alpha=1$ as in \eqref{eq:convdiv}.}
\label{fig:matfig8}
\end{figure}

\begin{figure}[htb]
\begin{center}
\includegraphics[width=0.48\textwidth]{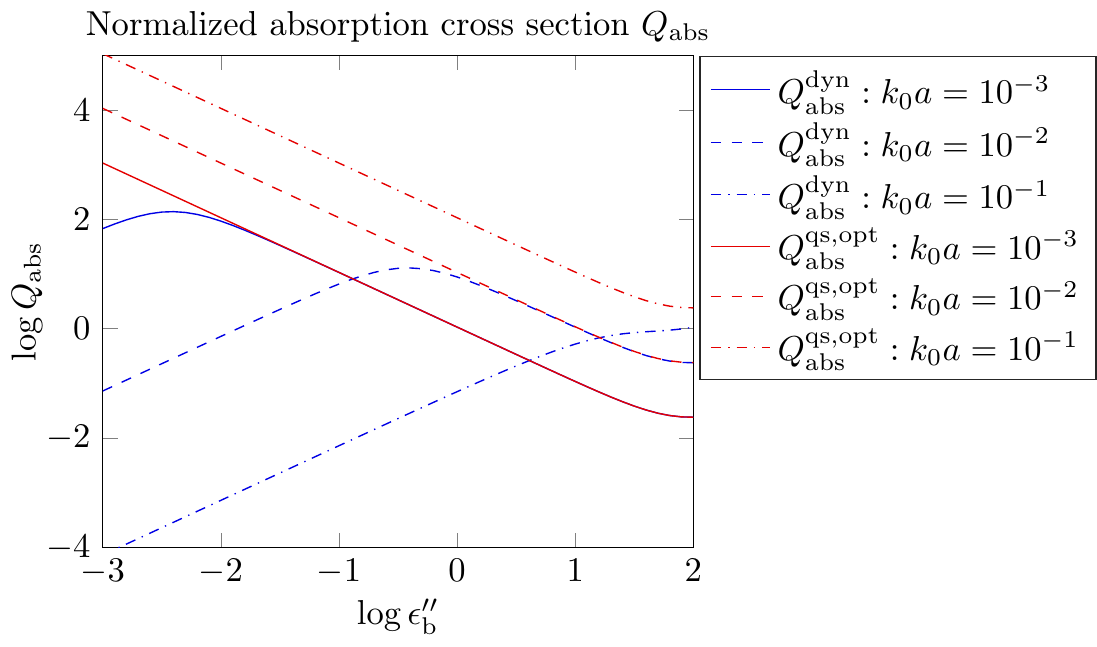}
\end{center}
\vspace{-5mm}
\caption{The quasistatic optimal normalized absorption cross section $Q_\mrm{abs}^\mrm{qs,opt}$ and the corresponding dynamic cross section $Q_\mrm{abs}^\mrm{dyn}$  
as functions of the background loss $\epsilon_\mrm{b}^{\prime\prime}$. Here, $\epsilon_\mrm{b}^{\prime}=80$, as with a biological tissue in the lower GHz region.}
\label{fig:matfig4b}
\end{figure}

In Figures \ref{fig:matfig3} through \ref{fig:matfig4b} we show the  normalized absorption cross section area $Q_\mrm{abs}$
and its behavior for $(k_0a,\epsilon_\mrm{b}^{\prime\prime})$ close to $(0,0)$.
Here, $Q_\mrm{abs}^\mrm{dyn}$ denotes the electrodynamic solution given by \eqref{eq:CabssphTMdipnorm} and $Q_\mrm{abs}^\mrm{qs,opt}$ corresponds to the optimal quasistatic solution
given by \eqref{eq:Cabssphere1o}, both of which are calculated for the background permittivity $\epsilon_\mrm{b}=\epsilon_\mrm{b}^{\prime}+\iu\epsilon_\mrm{b}^{\prime\prime}$
and the quasistatic optimal conjugate match $\epsilon=-2\epsilon_\mrm{b}^*$, yielding $\epsilon+2\epsilon_\mrm{b}=\iu 4\epsilon_\mrm{b}^{\prime\prime}$. 
In Figure \ref{fig:matfig3} we show also the normalized absorption cross section $Q_\mrm{abs}^\mrm{dyn,opt}$ corresponding
to the dynamic optimal solution \eqref{eq:epsilonospherethirdorder}, as well as the break points for the quasistatic approximation given by \eqref{eq:k0aoptasymptotics}.
As a comparison with the case with a lossless background $(\epsilon_\mrm{b}^{\prime\prime}=0)$, the optimal dipole absorption cross section 
$Q_\mrm{abs}^\mrm{dip}$ corresponding to \eqref{eq:Cabsdip} is also shown in Figure  \ref{fig:matfig3}.
In Figures \ref{fig:matfig3} through \ref{fig:matfig8} the background
is defined by $\epsilon_\mrm{b}^{\prime}=1$, and in Figure \ref{fig:matfig4b} the background corresponds to a saline water with $\epsilon_\mrm{b}^{\prime}=80$.

As can be seen in the Figures \ref{fig:matfig3} through \ref{fig:matfig8}, the normalized absorption cross section $Q_\mrm{abs}^\mrm{dyn}$
has a very subtle behavior for $(k_0a,\epsilon_\mrm{b}^{\prime\prime})$ close to the plasmonic singularity at $(0,0)$ where $\epsilon=-2$.
Note that $Q_\mrm{abs}^\mrm{dyn}\rightarrow 0$ as $k_0a\rightarrow 0$
 ($\epsilon_\mrm{b}^{\prime\prime}>0$ fixed) and $Q_\mrm{abs}^\mrm{dyn}\rightarrow 0$ as $\epsilon_\mrm{b}^{\prime\prime}\rightarrow 0$
 ($k_0a>0$ fixed), \cf also the approximate expression in \eqref{eq:CabssphTMdipnormapprox2}.
 However, as illustrated in Figure \ref{fig:matfig8},  $Q_\mrm{abs}^\mrm{dyn}$ is unbounded in
 any neighborhood where $(k_0a,\epsilon_\mrm{b}^{\prime\prime})$ is close to $(0,0)$ in accordance with the analysis given in Section \ref{sect:asymptoticanalys}.
 It should be noted that this behavior is perfectly consistent with the bound for a lossless background $C_\mrm{abs}\leq 3\lambda^2/8\pi$ given in \cite{Tretyakov2014}
 since $C_\mrm{abs}/a^2\leq (3/8\pi)/(a/\lambda)^2$ is unbounded as $a/\lambda$ approaches zero.
 
Figures  \ref{fig:matfig3}  through \ref{fig:matfig8} illustrate how the approximation $Q_\mrm{abs}^\mrm{qs,opt}\sim Q_\mrm{abs}^\mrm{dyn}$
becomes valid when $k_0a$ is sufficiently small  
at the same time as $\epsilon_\mrm{b}^{\prime\prime}$ is sufficiently large. 
In Figure \ref{fig:matfig3} is also illustrated that the break points given by \eqref{eq:k0aoptasymptotics} (or \eqref{eq:k0aoptasymptotics2})
give a very accurate estimate for the validity of the quasistatic approximation \eqref{eq:Cabssphere1o}.

The whole feasibility investigation regarding the quasistatic approximation above can 
readily be executed similarly for any other background permittivity $\epsilon_\mrm{b}^{\prime}$.
As \eg Figure \ref{fig:matfig4b} shows a comparison between  $Q_\mrm{abs}^\mrm{dyn}$ and $Q_\mrm{abs}^\mrm{qs,opt}$ 
with $\epsilon_\mrm{b}^{\prime}=80$ corresponding to the permittivity of biological tissue for frequencies 
in the lower GHz region, \cf \cite{Gabriel+etal1996b}. In this frequency region the corresponding
dielectric losses will be at least $\epsilon_\mrm{b}^{\prime\prime}>10$. Hence, the evaluation shown in Figure \ref{fig:matfig4b}
verifies that the previous investigations made in \cite{Nordebo+etal2017a,Dalarsson+etal2017a} are safely in the quasistatic regime
if \eg $k_0a<10^{-2}$, and which is certainly the case with cellular ($\mu$m) structures in the GHz frequency range.

\begin{figure}[htb]
\begin{center}
\includegraphics[width=0.48\textwidth]{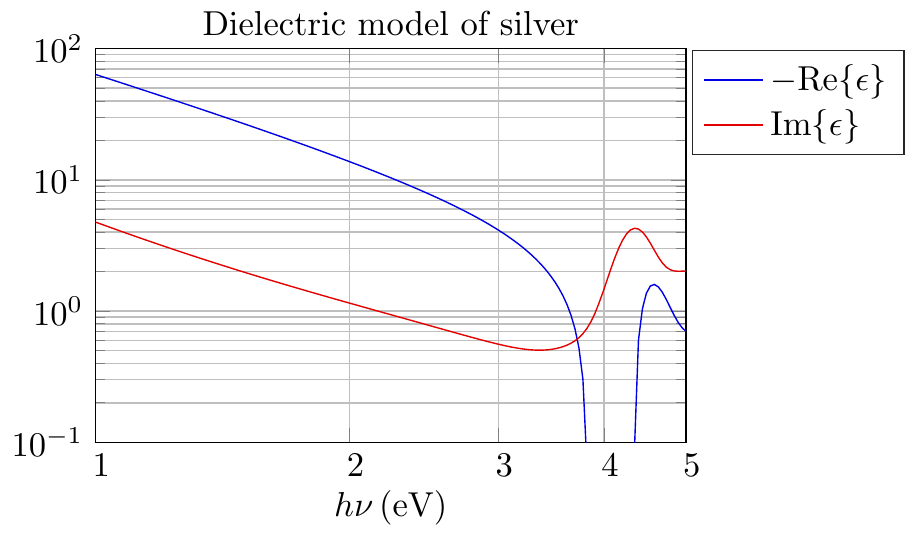}
\end{center}
\vspace{-5mm}
\caption{Dielectric function of silver according to the Brendel-Bormann model fitted to experimental data \cite{Rakic+etal1998}.}
\label{fig:matfig21}
\end{figure}

\begin{figure}[htb]
\begin{center}
\includegraphics[width=0.48\textwidth]{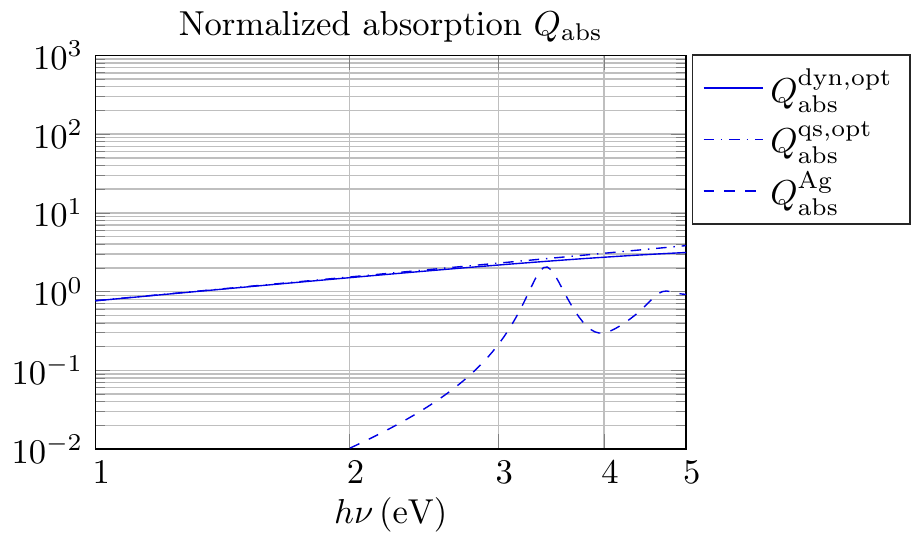}
\end{center}
\vspace{-5mm}
\caption{Normalized absorption cross sections for a sphere with radius  $a=10$\unit{nm} 
with $Q_\mrm{abs}^\mrm{dyn,opt}$ denoting the dynamic optimal absorption,  $Q_\mrm{abs}^\mrm{qs,opt}$ the quasistatic optimal absorption and 
$Q_{\mrm{abs}}^\mrm{Ag}$ the full dynamic dipole absorption corresponding to the Brendel-Bormann 
model for the dielectric function of silver \cite{Rakic+etal1998}, \cf also Fig.~\ref{fig:matfig21}.
Here, the background loss is given by $\epsilon_\mrm{b}=1+\iu 10^{-1}$.
}
\label{fig:matfig24}
\end{figure}

\begin{figure}[htb]
\begin{center}
\includegraphics[width=0.48\textwidth]{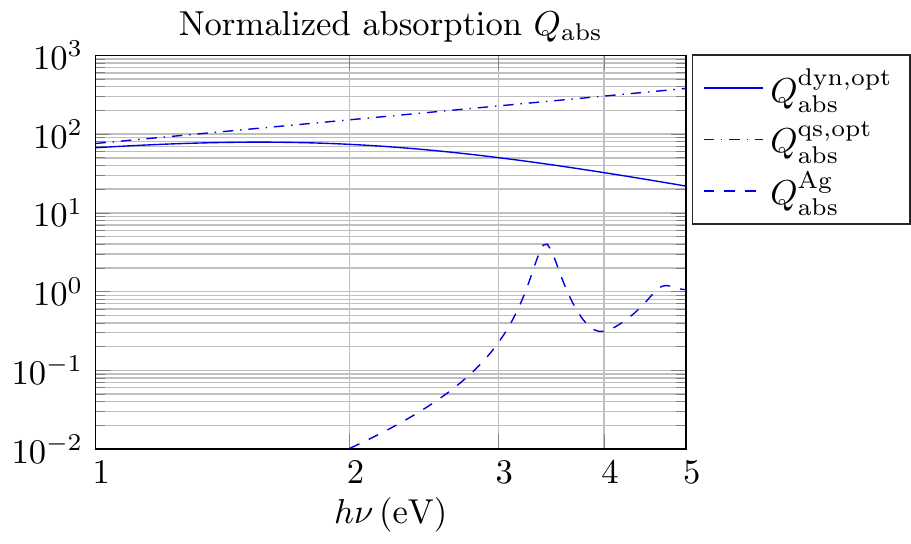}
\end{center}
\vspace{-5mm}
\caption{Same plot as in Fig.~\ref{fig:matfig24}, but here with the background loss given by $\epsilon_\mrm{b}=1+\iu 10^{-3}$.}
\label{fig:matfig24b}
\end{figure}

Finally, to illustrate the theory based on an application in plasmonics, we investigate the absorption in a silver 
nanosphere embedded in a slightly lossy medium.
Figure \ref{fig:matfig21} shows the dielectric function of silver according to the Brendel-Bormann (BB) model fitted to experimental data as presented in 
\cite[the dielectric model in Eq. (11) with parameter values from Table 1 and Table 3]{Rakic+etal1998}.
The frequency axis is given here in terms of the photon energy $h\nu$ in units of electron volts\unit{(eV)} where $h$ is Planck's constant and $\nu$ the frequency.
Figures \ref{fig:matfig24} and \ref{fig:matfig24b} present the normalized absorption cross section areas of a sphere with radius $a=10$\unit{nm} 
and where the quasistatic optimal $Q_\mrm{abs}^\mrm{qs,opt}$ is given by \eqref{eq:Cabssphere1o},  
the dynamic optimal $Q_\mrm{abs}^\mrm{dyn,opt}$ is given by \eqref{eq:CabssphTMdipnorm} 
together with \eqref{eq:epsilonospherethirdorder} and $Q_\mrm{abs}^\mrm{Ag}$ is given by \eqref{eq:CabssphTMdipnorm} together with the dielectric model of silver as illustrated
in Figure \ref{fig:matfig21}. The background medium is defined by $\epsilon_\mrm{b}=1+\iu\epsilon_\mrm{b}^{\prime\prime}$, where 
$\epsilon_\mrm{b}^{\prime\prime}=10^{-1}$ and $\epsilon_\mrm{b}^{\prime\prime}=10^{-3}$, respectively.
What is interesting to observe here is that the absorption of silver is in fact rather close to being optimal (the peak at $h\nu= 3.4$\unit{eV}) 
with the larger loss factor $\epsilon_\mrm{b}^{\prime\prime}=10^{-1}$,
and where the quasistatic and dynamic optimal solutions also agree rather well. With the smaller loss factor $\epsilon_\mrm{b}^{\prime\prime}=10^{-3}$,
the absorption is no longer close to being optimal and the quasistatic and dynamic optimal solutions diverge. 
Here, we have chosen the loss factors so that  $10^{-3}<\epsilon_\mrm{b,break}^{\prime\prime}<10^{-1}$ where
$\epsilon_\mrm{b,break}^{\prime\prime}=0.03$ corresponds to the break point defined by \eqref{eq:k0aoptasymptotics2} for resonance at $h\nu=3.4$\unit{eV}.

\section{Summary and conclusions}

It has been demonstrated that the maximal dipole absorption of a small dielectric or conducting sphere is 
unbounded under the quasistatic approximation if the losses in the surrounding medium can be made arbitrarily small. 
This deviation has been rectified by using the general Mie theory and an asymptotic analysis to give explicit formulas to assess the validity of the quasistatic approximation.
In particular, it turns out that the quasistatic theory is valid provided that the electrical size of the sphere is sufficiently small
at the same time as the exterior losses are sufficiently large. Moreover, it has been shown that the optimal normalized absorption cross section area of 
the small dielectric sphere has a very subtle limiting behavior, and is in fact unbounded when both the electrical size as well as the exterior losses tend to zero.
Finally,  an improved asymptotic formula based on full dynamics has been given for the optimal plasmonic dipole absorption of the sphere, which is 
valid for small spheres as well as for small losses. Numerical examples have been included to illustrate the asymptotic results.

\begin{acknowledgments}
This work has been partly supported by the Swedish Foundation for Strategic Research (SSF) under the programme 
Applied Mathematics and the project Complex Analysis and Convex Optimization for EM Design.
\end{acknowledgments}

\appendix
\section{Spherical vector waves}\label{sect:spherical}
\subsection{Definition of spherical vector waves}\label{sect:sphericaldef}
In a source-free homogeneous and isotropic medium the electromagnetic field can be expanded in spherical vector waves as
\begin{equation}\label{eq:Esphdef}
\bm{E}(\bm{r})=\sum_{l,m,\tau}a_{\tau ml}{\bf v}_{\tau ml}(k\bm{r})+b_{\tau ml}{\bf u}_{\tau ml}(k\bm{r}),  
\end{equation}
and
\begin{equation}\label{eq:Hsphdef}
\bm{H}(\bm{r})=\frac{1}{\iu\eta_0\eta}
\sum_{l,m,\tau}a_{\tau ml}{\bf v}_{\bar{\tau} ml}(k\bm{r})+b_{\tau ml}{\bf u}_{\bar{\tau} ml}(k\bm{r}),  
\end{equation}
where ${\bf v}_{\tau ml}(k\bm{r})$ and ${\bf u}_{\tau ml}(k\bm{r})$ are the regular and the outgoing spherical vector waves, respectively, 
and $a_{\tau ml}$ and $b_{\tau ml}$ the corresponding multipole coefficients,
see \eg \cite{Arfken+Weber2001,Jackson1999,Newton2002,Kristensson2016}.
Here, $l=1,2,\ldots,$ is the multipole order, $m=-l,\ldots,l$ the azimuthal index and $\tau=1,2$, where
$\tau=1$ indicates a transverse electric (\textrm{TE}) magnetic multipole and $\tau=2$ a transverse magnetic (\textrm{TM}) electric multipole,
and $\bar{\tau}$ denotes the dual index, \ie $\bar{1}=2$ and $\bar{2}=1$.

The solenoidal (source-free) regular spherical vector waves are defined here by
\begin{multline}\label{eq:v1def}
\displaystyle{\bf v}_{1 ml}(k{\bm{r}})  =   \frac{1}{\sqrt{l(l+1)}}\nabla\times({\bm{r}}\mrm{j}_l(kr)\mrm{Y}_{ml}(\hat{\bm{r}})) \\
 =   \mrm{j}_l(kr){\bf A}_{1 ml}(\hat{\bm{r}}), 
\end{multline}
and
\begin{multline}\label{eq:v2def}
{\bf v}_{2 ml}(k\bm{r})   =   \displaystyle \frac{1}{k}\nabla\times{\bf v}_{1 ml}(k\bm{r}) \\
  =\displaystyle\frac{(kr\mrm{j}_l(kr))^{\prime}}{kr}{\bf A}_{2 ml}(\hat{\bm{r}})+\sqrt{l(l+1)}\frac{\mrm{j}_l(kr)}{kr}{\bf A}_{3 ml}(\hat{\bm{r}}),
\end{multline}
where $\mrm{Y}_{ml}(\hat{\bm{r}})$ are the spherical harmonics, ${\bf A}_{\tau ml}(\hat{\bm{r}})$ the vector spherical harmonics and $\mrm{j}_l(x)$ the spherical Bessel functions of order $l$,
\cf \cite{Arfken+Weber2001,Jackson1999,Newton2002,Olver+etal2010,Kristensson2016}. 
Here, $(\cdot)^\prime$ denotes a differentiation with respect to the argument of the spherical Bessel function.
The outgoing (radiating) vector spherical waves ${\bf u}_{\tau ml}(k{\bm{r}})$ are obtained by replacing
the regular spherical Bessel functions $\mrm{j}_l(x)$ above for the spherical Hankel functions of the first kind, $\mrm{h}_l^{(1)}(x)$, 
see \cite{Jackson1999,Olver+etal2010,Kristensson2016}.

The vector spherical harmonics ${\bf A}_{\tau ml}(\hat{\bm{r}})$ are given by
\begin{equation}\label{eq:Adef}
\left\{\begin{array}{l}
{\bf A}_{1ml}(\hat{\bm{r}})  =   \displaystyle\frac{1}{\sqrt{l(l+1)}}\nabla\times\left( \bm{r}\mrm{Y}_{ml}(\hat{\bm{r}}) \right),   \vspace{0.2cm} \\
{\bf A}_{2ml}(\hat{\bm{r}})  =  \hat{\bm{r}}\times{\bf A}_{1ml}(\hat{\bm{r}}),  \vspace{0.2cm} \\
{\bf A}_{3ml}(\hat{\bm{r}}) = \hat{\bm{r}}\mrm{Y}_{ml}(\hat{\bm{r}}), 
\end{array}\right.
\end{equation}
where $\tau=1,2,3$, and where the spherical harmonics $\mrm{Y}_{ml}(\hat{\bm{r}})$ are given by
\begin{eqnarray}
\mrm{Y}_{ml}(\hat{\bm{r}})=(-1)^m\sqrt{\frac{2l+1}{4\pi}}\sqrt{\frac{(l-m)!}{(l+m)!}}\mrm{P}_{l}^m(\cos\theta)\eu^{{\rm i}m\phi}, \nonumber
\end{eqnarray}
and where $\mrm{P}_{l}^m(x)$ are the associated Legendre functions \cite{Arfken+Weber2001,Jackson1999,Olver+etal2010}.
The associated Legendre functions can be obtained from
\begin{equation}
\mrm{P}_l^m(\cos\theta)=(-1)^m(\sin\theta)^m\frac{\mrm{d}^m}{\mrm{d}(\cos\theta)^m}\mrm{P}_l(\cos\theta),   
\end{equation}
where $\mrm{P}_l(x)$ are the Legendre polynomials of order $l$ and $0\leq m\leq l$,  see \cite{Arfken+Weber2001,Jackson1999,Olver+etal2010}.
Important symmetry properties are $\mrm{P}_l^{-m}(x)=(-1)^m\frac{(l-m)!}{(l+m)!}\mrm{P}_l^m(x)$ and 
$\mrm{Y}_{-m,l}(\theta,\phi)=(-1)^m\mrm{Y}_{ml}^*(\theta,\phi)$ where $m\geq 0$. Hence, the vector spherical harmonics satisfy the symmetry 
${\bf A}_{\tau, -m,l}(\hat{\bm{r}})=(-1)^m{\bf A}^{*}_{\tau ml}(\hat{\bm{r}})$.
The vector spherical harmonics are orthonormal on the unit sphere, and hence
\begin{equation}\label{eq:Aorthonormal}
\int_{\Omega_0}{\bf A}_{\tau ml}^*(\hat{\bm{r}})\cdot{\bf A}_{\tau^\prime m^\prime l^\prime }(\hat{\bm{r}})\mrm{d}\Omega
=\delta_{\tau\tau^\prime}\delta_{mm^\prime}\delta_{ll^\prime},
\end{equation}
where $\Omega_0$ denotes the unit sphere and $\mrm{d}\Omega=\sin\theta\mrm{d}\theta\mrm{d}\phi$. 

\subsection{First Lommel integral for spherical Bessel functions with complex valued argument}\label{sect:Lommel}
Let ${\rm s}_l(kr)$ denote an arbitrary linear combination of spherical Bessel and Hankel functions.
Based on the first Lommel integral for cylinder functions, \cf \cite[Eq.~(10.22.4) on p.~241]{Olver+etal2010} and \cite[Eq.~(8) on p.~134]{Watson1995},
the following indefinite Lommel integral can be derived for spherical Bessel functions
\begin{eqnarray}\label{eq:firstLommel}
\int\left|{\rm s}_l(kr)\right|^2r^2\mrm{d} r
=r^2\frac{ \Im\!\left\{k{\rm s}_{l+1}(kr) {\rm s}_{l}^*(kr) \right\}}{\Im\!\left\{k^2\right\}},
\end{eqnarray}
where $k$ is complex valued ($k\neq k^*$), \cf \cite[Eq.~(A.15) on p.~11]{Nordebo+etal2017a}.
Furthermore, by using the recursive relationships
\begin{equation}\label{eq:recursive}
\left\{\begin{array}{l}
\displaystyle \frac{\mrm{s}_l(kr)}{kr}=\frac{1}{2l+1}\left(\mrm{s}_{l-1}(kr)+\mrm{s}_{l+1}(kr) \right), \vspace{0.2cm}  \\
\displaystyle \mrm{s}_l^\prime(kr)=\frac{1}{2l+1}\left(l\mrm{s}_{l-1}(kr)-(l+1)\mrm{s}_{l+1}(kr) \right), 
\end{array}\right.
\end{equation}
where $l=1,2,\ldots$, \cf \cite{Olver+etal2010}, it can be shown that
\begin{multline}\label{eq:recursiveLommel}
\int\left(\left|\frac{\mrm{s}_{l}(kr)}{kr}+\mrm{s}_{l}^\prime(kr)\right|^2 
 +l(l+1)  \left|\frac{\mrm{s}_{l}(kr)}{kr}\right|^2\right)r^2\mrm{d} r  \\
 =\frac{1}{2l+1}\int \left( (l+1)\left|{\rm s}_{l-1}(kr)\right|^2+l\left|{\rm s}_{l+1}(kr)\right|^2\right)r^2\mrm{d} r, 
\end{multline}
see also \eg \cite[Eq.~(17) on p.~411]{Marengo+Devaney1999} and \cite[Eqs.~(36) and (47) on pp.~2359--2360]{Nordebo+Gustafsson2006}.

\subsection{Orthogonality of the regular spherical waves}\label{sect:sphericalorth}
Due to the orthonormality of the vector spherical harmonics \eqref{eq:Aorthonormal},
the regular spherical vector waves are orthogonal over the unit sphere with
\begin{equation}\label{eq:vorthogonal1}
\displaystyle\int_{\Omega_0}{\bf v}_{\tau ml}^*(k{\bm{r}})\cdot{\bf v}_{\tau^\prime m^\prime l^\prime}(k{\bm{r}})\mrm{d} \Omega 
=\displaystyle\delta_{\tau\tau^\prime}\delta_{mm^\prime}\delta_{ll^\prime}S_{\tau l}(k,r), 
\end{equation}
where
\begin{multline}\label{eq:Stauldef}
S_{\tau l}(k,r)=\displaystyle\int_{\Omega_0}|{\bf v}_{\tau ml}(k\bm{r})|^2\mrm{d}\Omega  \\
 =\left\{\begin{array}{ll}
\displaystyle  \left|\mrm{j}_{l}(kr)\right|^2 & \textrm{for}\  \tau=1,  \vspace{0.2cm} \\
\displaystyle  \left|\frac{\mrm{j}_{l}(kr)}{kr}+\mrm{j}_{l}^\prime(kr)\right|^2+l(l+1) \left|\frac{\mrm{j}_{l}(kr)}{kr}\right|^2  & \textrm{for}\  \tau=2. 
\end{array}\right.
\end{multline}
As a consequence, the regular spherical vector waves are also orthogonal over a spherical volume $V_{a}$ with radius $a$ yielding
\begin{equation}\label{eq:vorthogonal2}
\displaystyle\int_{V_{a}}{\bf v}_{\tau ml}^*(k{\bm{r}})\cdot{\bf v}_{\tau^\prime m^\prime l^\prime}(k{\bm{r}})\mrm{d} v 
=\displaystyle\delta_{\tau\tau^\prime}\delta_{mm^\prime}\delta_{ll^\prime}W_{\tau l}(k,a), 
\end{equation}
where
\begin{equation}\label{eq:Wtauldef}
W_{\tau l}(k,a)=\int_{V_{a}}\left|{\bf v}_{\tau ml}(k\bm{r})\right|^2\mrm{d} v 
=\int_{0}^{a}S_{\tau l}(k,r)r^2\mrm{d} r, 
\end{equation}
where $\mrm{d} v=r^2\mrm{d}\Omega\mrm{d} r$ and $\tau=1,2$.

For complex valued arguments $k$, $W_{1l}(k,a)$ is obtained from \eqref{eq:firstLommel} as
\begin{equation}\label{eq:W1ldef}
W_{1l}(k,a)=\int_{0}^{a}\left|\mrm{j}_{l}(kr)\right|^2 r^2\mrm{d} r 
=\frac{a^2 \Im\!\left\{k\mrm{j}_{l+1}(ka) \mrm{j}_{l}^*(ka) \right\}}{\Im\!\left\{k^2\right\}}, 
\end{equation}
and from \eqref{eq:recursiveLommel} follows that
\begin{multline}\label{eq:W2ldef}
W_{2 l}(k,a) \\
=\int_{0}^{a}\left(\left|\frac{\mrm{j}_{l}(kr)}{kr}+\mrm{j}_{l}^\prime(kr)\right|^2 
 +l(l+1)  \left|\frac{\mrm{j}_{l}(kr)}{kr}\right|^2\right)r^2\mrm{d} r  \\
 =\frac{1}{2l+1}\left((l+1)W_{1,l-1}(k,a)+lW_{1,l+1}(k,a) \right). 
\end{multline}


\end{document}